\documentclass[aps,prl,twocolumn,showpacs,preprintnumbers,amsmath,amssymb,superscriptaddress]{revtex4}%

\usepackage{graphicx}%
\usepackage{dcolumn}
\usepackage{amsmath}
\usepackage{color}

\begin{document}

\preprint{PREPRINT (\today)}

\title{ Evidence for competition between the superconducting and the pseudogap state in (BiPb)$_2$(SrLa)$_2$CuO$_{6+\delta}$ from muon-spin rotation experiments}

\author{R.~Khasanov}
 \email{rustem.khasanov@psi.ch}
 \affiliation{Laboratory for Muon Spin Spectroscopy, Paul Scherrer
Institut, CH-5232 Villigen PSI, Switzerland}
 \affiliation{Physik-Institut der Universit\"{a}t Z\"{u}rich,
Winterthurerstrasse 190, CH-8057 Z\"urich, Switzerland}

\author{Takeshi~Kondo}
 \affiliation{Ames Laboratory and Department of Physics and Astronomy, Iowa State University, Ames, IA~50011, USA}
 \affiliation{Department of Crystalline Materials Science, Nagoya University, Nagoya 464-8603, Japan}
\author{S.~Str\"assle}
 \affiliation{Physik-Institut der Universit\"{a}t Z\"{u}rich,
Winterthurerstrasse 190, CH-8057 Z\"urich, Switzerland}
\author{D.O.G.~Heron}
 \affiliation{School of Physics and Astronomy, University of St.~Andrews, Fife,
KY16 9SS, UK}
\author{A.~Kaminski}
 \affiliation{Ames Laboratory and Department of Physics and Astronomy, Iowa
State University, Ames, IA~50011, USA}
\author{H.~Keller}
 \affiliation{Physik-Institut der Universit\"{a}t Z\"{u}rich,
Winterthurerstrasse 190, CH-8057 Z\"urich, Switzerland}
\author{S.L.~Lee}
 \affiliation{School of Physics and Astronomy, University of St.~Andrews, Fife,
KY16 9SS, UK}
\author{Tsunehiro Takeuchi}
\affiliation{Department of Crystalline Materials Science, Nagoya
University, Nagoya 464-8603, Japan} \affiliation{EcoTopia Science
Institute, Nagoya University, Nagoya 464-8603, Japan}

\begin{abstract}
The in-plane magnetic penetration depth $\lambda_{ab}$  in
optimally doped (BiPb)$_2$(SrLa)$_2$CuO$_{6+\delta}$ (OP Bi2201)
was studied by means of muon-spin rotation. The measurements
of $\lambda_{ab}^{-2}(T)$ are inconsistent with a simple model of
a $d-$wave order parameter and a uniform quasiparticle weight
around the Fermi surface. The  data are well described
assuming the angular gap symmetry obtained in
ARPES experiments
[Phys.~Rev.~Lett  {\bf 98}, 267004 (2007)],
where it was shown that the superconducting gap in OP Bi2201
exists only in segments of the Fermi surface near the nodes.
We find that the remaining parts of the Fermi surface, which are
strongly affected by the pseudogap state, do not contribute
significantly to the superconducting condensate. Our data provide
evidence that high temperature superconductivity and pseudogap behavior in
cuprates are competing phenomena.
\end{abstract}

\maketitle


The relevance of the pseudogap phenomenon for superconductivity 
is an important open issue in the physics of high-temperature
cuprate superconductors (HTS's). There are two main scenarios to
be considered. In the first, the so-called "precursor scenario``, the
Cooper pairs are already formed at $T^\ast$,  the temperature at
which the pseudogap opens first, but long-range phase coherence is not
established until the sample is cooled below the superconducting
transition temperature $T_c$. In the second, the so-called
''two-gap`` scenario, the superconducting and the pseudogap state are
not directly related with each other, and may even compete.
Within this scenario the gaps in $k-$space, existing near the
nodes and in the antinodal region of the Fermi surface, are due to the superconducting
and the pseudogap states, respectively. This scenario gained support
due to a number of recent experiments
\cite{LeTacon06,Tanaka06,Kondo07,Lee07,Guyard08} which revealed
that the antinodal gap remains unaffected as the temperature
changes across $T_c$, and generally its magnitude increases
significantly in the underdoped region, where $T_c$ decreases. In
contrast, the gap near the nodes scales with $T_c$ and obeys a
well defined BCS temperature dependence \cite{Lee07}. This
interpretation also agrees with recent results  from
scanning-tunneling-microscopy experiments of Hanaguri {\it et al.}
\cite{Hanaguri07}, suggesting that the incoherent antinodal
states are not responsible for the formation of phase-coherent Cooper
pairs.  Consequently, superconductivity is caused by the
coherent part of the Fermi surface near the nodes.

Measurements of the magnetic penetration depth  $\lambda$ can be
used to distinguish between the above described scenarios. The
temperature dependence of $\lambda$ is uniquely determined by the
absolute maximum value of the superconducting energy gap and its
angular and temperature dependence. In addition, within the
London model $\lambda^{-2}$ is proportional to the superfluid
density via $\lambda^{-2}\propto\rho_s\propto n_s/m^\ast$
and, in case where the supercarrier mass $m^\ast$ is known,
gives information on the supercarrier density  $n_s$.

Here we report on a study of the in-plane magnetic penetration
depth $\lambda_{ab}$ in optimally doped
(BiPb)$_2$(SrLa)$_2$CuO$_{6+\delta}$ (OP Bi2201). The angular
dependence of the energy gap in similar OP Bi2201 samples was recently  studied by Kondo {\it et al.} \cite{Kondo07} by
means of angular-resolved photoemission (ARPES), where the
observation of two spectral gaps that dominate different regions
of the Fermi surface is reported. Our results reveal that
$\lambda^{-2}_{ab}(T)$ is inconsistent with a model in which both
of these spectral gaps are related to superconductivity, as well as with a
superconducting gap of $d-$wave symmetry developed within the
whole Fermi surface.
Good agreement with the ARPES data was obtained within a model which
assumes that the pseudogap affects the spectral density of the antinodal 
quasiparticles. Consequently,  only carriers close to the
nodes contribute to the superfluid density, while the weight of the coherent quasiparticle
near the antinodes is negligible. This statement is also
supported by comparing the zero-temperature value of $\lambda_{ab}^{-2}(0)$
for OP Bi2201 studied here with those of other OP HTS's, such as
Ca$_{2-x}$Na$_x$CuO$_2$Cl$_2$ (OP Na-CCOC)
\cite{Khasanov07_NaCCOC}  and La$_{2-x}$Sr$_{x}$CuO$_4$ (OP La214)
\cite{Panagopoulos99}, having similar transition temperatures. It
was observed that in superconductors where the superconducting gap
is developed only close to the nodes (OP Bi2201 and OP Na-CCOC) the
superfluid density is more than 50\% smaller than in OP La214
where the $d-$wave superconducting gap is detected on the whole
Fermi surface \cite{Shi07}.


Details on the sample  preparation for OP Bi2201
single crystals
can be found elsewhere \cite{Kondo04Kondo05}. The values of $T_c$
and the width of the superconducting transition, as determined
from magnetization measurements, are $\simeq$35~K and $\simeq$3~K,
respectively.
The transverse-field muon-spin  rotation (TF-$\mu$SR)  experiments
were performed at the $\pi$M3 beam line at the Paul Scherrer
Institute (Villigen, Switzerland). Two OP
Bi2201 single crystals  with an approximate size of 4$\times$2$\times$0.1~mm$^3$
were mounted on a holder specially designed to perform $\mu$SR
experiments on thin single crystalline samples. The
sample was field cooled from above $T_c$ to 1.6~K in a series of
fields ranging from 5~mT to 640~mT. The magnetic field was applied
parallel to the crystallographic $c$ axis and transverse to the
muon-spin polarization.

\begin{figure}[htb]
\includegraphics[width=0.9\linewidth]{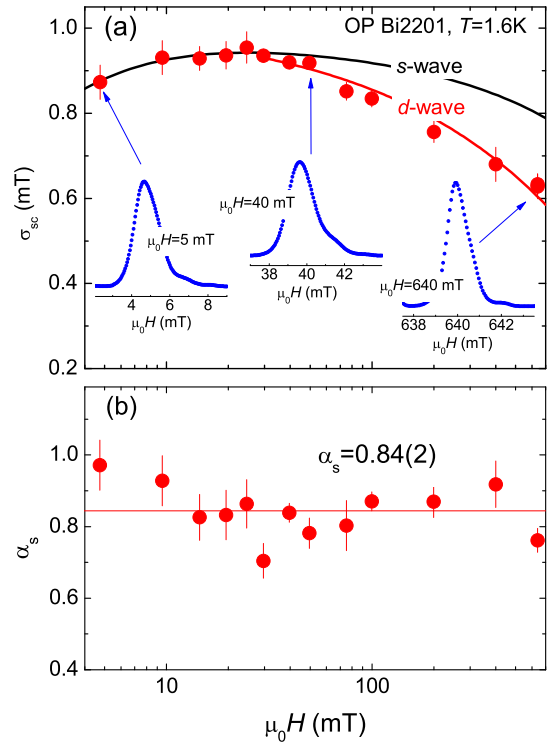}
\caption{(Color online) (a) Dependence of $\sigma_{sc}$ of OP Bi2201 on the
applied magnetic field measured at $T=1.6$~K. The black  solid
line corresponds to $\sigma_{sc}(H)$ obtained by using the
numerical calculations of Brandt \cite{Brandt03}
($\lambda=360$~nm, $\kappa=140$) for a superconductor with an
isotropic energy gap. The solid red line represents
$\sigma_{sc}(H)$ expected in case of a $d-$wave superconductor.
The blue dotted curves show the local magnetic field distribution
$P(B)$ calculated by means of the maximum entropy Fourier-transform technique at $T=1.6$~K and $\mu_0H=5$~mT, 40~mT, and
640~mT. (b) Field dependence of the skewness parameter
$\alpha_{s}$. The solid line is the average value $\alpha_s=0.84(2)$.}
 \label{fig:sigma_vs_H}
\end{figure}

In the TF geometry the  local  magnetic  field distribution $P(B)$
inside a superconductor in the mixed state, probed by means of
$\mu$SR, is determined by the coherence length $\xi$,
and the penetration depth $\lambda$. In extreme type-II
superconductors ($\lambda\gg\xi$) the $P(B)$ distribution is
almost independent of $\xi$ and the second moment of $P(B)$
becomes proportional to $1/\lambda^4$ \cite{Brandt03}.
To describe the asymmetric $P(B)$ (see
Fig.~\ref{fig:sigma_vs_H}), the $\mu$SR time spectra  were
analyzed by using a two-component Gaussian expression
\cite{Khasanov06a}. The second moment of $P(B)$ was further
obtained as \cite{Khasanov06a}:
\begin{equation}
\langle \Delta B^{2}\rangle=\frac{\sigma^2}{\gamma^2_\mu}=
\sum_{i=1}^2{A_i \over A_1+A_2} \left[
\frac{\sigma_i^2}{\gamma^2_\mu} +\left(B_i-
\frac{A_iB_i}{A_1+A_2}\right)^2 \right] .
\label{eq:dB}
\end{equation}
Here $A_i$, $\sigma_i$, and $B_i$ are the asymmetry,  the
relaxation rate, and the mean field of the $i-$th component, and
$\gamma_\mu = 2\pi\times135.5342$~MHz/T is the muon gyromagnetic
ratio. The analysis was simplified to a single Gaussian lineshape in
the case when the  two-Gaussian and the one-Gaussian fits result in
comparable $\chi^2$. The superconducting part 
of the second moment $\sigma_{sc}^2$ was obtained by subtracting the
contribution of the nuclear moments $\sigma_{nm}$ measured at $T>T_c$
as $\sigma_{sc}^2=\sigma^2 - \sigma_{nm}^2$ \cite{Khasanov06a}.
Since the magnetic field was applied along the crystallographic
$c$ axis, our experiments provide direct information on $\lambda_{ab}$.

Fig.~\ref{fig:sigma_vs_H}a shows the dependence of
$\sigma_{sc}$ on the applied magnetic field measured after field
cooling the OP Bi2201 sample from $T>T_c$ down to 1.6~K. The
$P(B)$ distributions were calculated using the maximum entropy
Fourier-transform technique for $\mu_0H=5$~mT, 40~mT, and 640~mT
(see Fig.~\ref{fig:sigma_vs_H}a).
In the whole range  of fields ($5$~mT$\leq\mu_0H\leq640$~mT)
$P(B)$ is asymmetric. The asymmetric shape of $P(B)$ is generally
described in terms of the so-called skewness parameter
$\alpha_{s}=\langle \Delta B^{3}\rangle^{1/3}/\langle \Delta
B^{2}\rangle^{1/2}$ [$\langle \Delta B^{n}\rangle$ is the $n-$th
central moment of $P(B)$]. $\alpha_{s}$ is a dimensionless
measure of the asymmetry of the lineshape, the variation of which
reflects underlying changes in the vortex structure \cite{Lee93}.
In the limit $\kappa \gg 1$ and for realistic
measuring conditions
$\alpha_{s}\simeq1.2$ for an ideal triangular vortex lattice (VL).  It is very sensitive to structural
changes of the VL which can occur as a function of 
temperature and/or magnetic field \cite{Lee93,Aegerter98}.
Fig.~\ref{fig:sigma_vs_H}b implies that in OP Bi2201
$\alpha_{s}(H)$ is almost constant [$\alpha_s=0.84(2)$] and is smaller than the
expected value of 1.2, which is probably caused by distortions of
the VL due to pinning effects. It is known that Pb substitution
in double-layer Bi2212 HTS's enhances pinning quite
substantially \cite{Musolino03}.

It should be noted here that addition of Pb does not change $T_c$ and the
in-plane superfluid density $\rho_s\propto\lambda^{-2}_{ab}$
\cite{Russo07}, but makes OP Bi2201 {\it more} 3-dimensional. To estimate the anisotropy coefficient
$\gamma_{c,ab}=\lambda_c/\lambda_{ab}$ ($\lambda_c$ is the
$c-$axis component of the penetration depth) we performed torque
magnetization experiment on one of the crystals studied
\cite{Weyeneth08}. A value of $\gamma_{c,ab}\simeq20$ was found,
which is more than 10 times smaller than
$\gamma_{c,ab}\simeq200-400$ obtained on OP Bi2201 without Pb by
Kawamata {\it et al.} \cite{Kawamata99}.

Figure~\ref{fig:sigma_vs_H} indicates that the magnetic field dependence of
$\sigma_{sc}$ is not monotonic: with increasing
field $\sigma_{sc}$  goes through the broad maximum at around 20~mT.
The black solid line in Fig.~\ref{fig:sigma_vs_H}a, calculated within the model of Brandt \cite{Brandt03},
corresponds to $\sigma_{sc}(H)$ for an isotropic $s-$wave
superconductor with $\lambda=360$~nm and
$\kappa=\lambda/\xi\simeq140$ ($\xi\simeq2.6$~nm was obtained from
the value of the second critical field $\mu_0H_{c2}(0)\simeq50$~T
\cite{Wang03}). From Fig.~\ref{fig:sigma_vs_H}a we conclude that the
experimental $\sigma_{sc}(H)$ depends much stronger on the
magnetic field than expected for a fully gaped
$s-$wave superconductor. As shown by Amin {\it et al.}
\cite{Amin00} for a superconductor with nodes in the energy gap a field dependent correction to $\rho_s$ arises
from its nonlocal and nonlinear response to an applied magnetic
field. The solid
red line represents the result of the fit by means of the relation:
\begin{equation}
\frac{\rho_s(H)}{\rho_s(H=0)}=
\frac{\sigma_{sc}(H)}{\sigma_{sc}(H=0)}= 1-K \sqrt{H},
 \label{eq:lambda_vs_H}
\end{equation}
which takes the nonlinear correction to $\rho_s$ for
a superconductor with a $d-$wave energy gap  into account  \cite{Vekhter99}. Here
the parameter $K$ depends on the strength of the nonlinear effect.
Since Eq.~(\ref{eq:lambda_vs_H}) is valid for intermediate
fields $H_{c1}\ll H\ll H_{c2}$ ($H_{c1}$ is the first critical
field) only the data points above 40~mT were considered in the
analysis.

\begin{figure}[htb]
\includegraphics[width=0.9\linewidth]{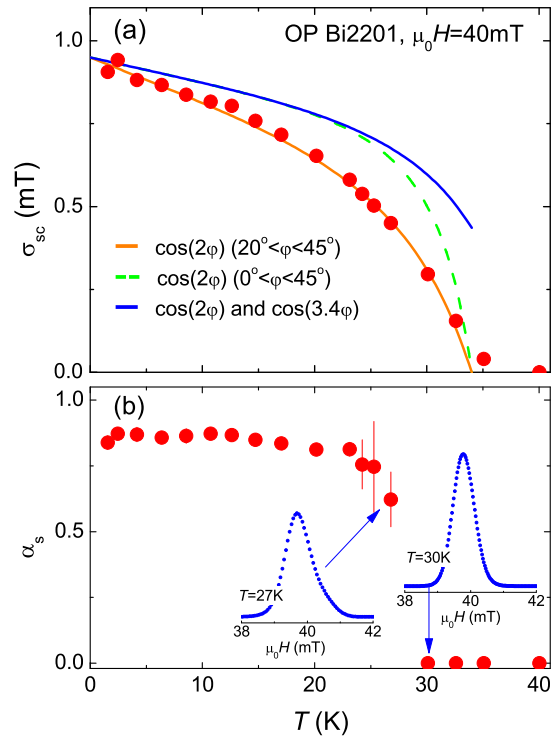}
\caption{(Color online) (a) Dependence of
$\sigma_{sc}$ of
OP Bi2201 on $T$ measured at $\mu_0H=40$~mT. Lines
represent the theoretical $\sigma_{sc}(T)$ curves obtained by assuming different
symmetries of the superconducting energy gap (see Fig.~\ref{fig:gap}a). (b) Dependence of the skewness parameter
$\alpha_{s}$ on $T$. The blue dotted curves represent $P(B)$
distributions below ($T=27$~K) and above ($T=30$~K) the
VL melting temperature. }
 \label{fig:sigma_vs_T}
\end{figure}

We now discuss the $T$ dependence of $\sigma_{sc}$.
Figure~\ref{fig:sigma_vs_T} displays $\sigma_{sc}(T)$ measured at
$\mu_0H=40$~mT. Below 20~K $\sigma_{sc}$ is linear in $T$ as
expected for a superconductor with nodes in the gap, consistent with the conclusion drawn from the analysis of the $\sigma_{sc}(H)$ data (see
discussion above and Fig.~\ref{fig:sigma_vs_H}).
To ensure that $\sigma_{sc}(T)$ is determined primarily by the
variance of the magnetic field within the VL we plot in
Fig.~\ref{fig:sigma_vs_T}b the corresponding
$\alpha_{s}(T)$. It is constant from 1.6~K to $\simeq27$~K and drops to zero at $T\simeq30$~K, where $P(B)$ becomes fully symmetric. A similarly sharp change of $\alpha_{s}$ with
temperature was observed in Bi2212 and was explained by VL
melting \cite{Lee93,Aegerter98}.
Correspondingly, we conclude that for temperatures $0<T\lesssim30$~K the $T$ variation of $\sigma_{sc}$ reflects the {\it intrinsic} behavior of the in-plane
magnetic penetration depth $\lambda_{ab}$.

\begin{figure}[htb]
\includegraphics[width=0.9\linewidth]{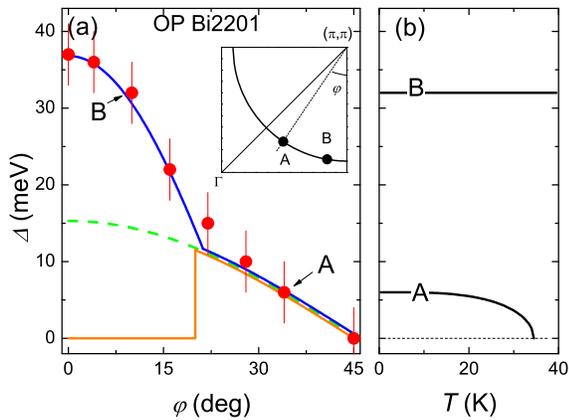}
\caption{(Color online) (a) Angular dependence of the energy gap of OP Bi2201
obtained in ARPES experiments \cite{Kondo07}. Lines represent the various models of the gap symmetries
used to analyze the experimental $\sigma_{sc}(T)$ data (see
Fig.~\ref{fig:sigma_vs_T}a). The inset shows schematically a
part of the Fermi surface. The points ''A`` and ''B`` are close to
the nodal ($\varphi\sim45^{\rm o}$) and the antinodal ($\varphi\sim0^{\rm o}$)
region, respectively.  (b) Temperature dependence of the energy
gap in the nodal (curve A) and  the antinodal (curve B) regions. }
 \label{fig:gap}
\end{figure}

The $T$ dependence of $\sigma_{sc}$ was analyzed by
assuming that the angular dependence of the energy gap in OP
Bi2201 is similar to the one from recent ARPES experiments
\cite{Kondo07} (see
Fig.~\ref{fig:gap}a). In analogy with
Refs.~\onlinecite{Kondo07} and \onlinecite{Lee07} it was also
assumed that the energy gap in the nodal region changes with
temperature in accordance with the weak-coupling BCS prediction
$\widetilde{\Delta}(T/T_c)=\tanh\{1.82[1.018(T_c/T-1)^{0.51}] \}$
\cite{Carrington03}, while the one near the antinodes is $T$ independent (see the corresponding lines
''A`` and ''B`` in Fig.~\ref{fig:gap}b). The following cases
were considered: (I) a monotonic $d-$wave gap 
$\Delta(T,\varphi)=15\ {\rm
meV}\cdot\cos(2\varphi)\widetilde{\Delta}(T/T_c)$ (green dashed
line); (II) a monotonic $d-$wave gap with  suppressed
quasiparticle weight in the antinodal region (solid orange line);
(III) an analytical function, which follows the monotonic $d-$wave
$15\ {\rm meV}\cdot\cos(2\varphi)\widetilde{\Delta}(T/T_c)$ in the
nodal region and changes to a $36\ {\rm meV}\cdot\cos(3.4\varphi)$
behavior close to the antinodes (solid blue line).
The $T$ dependence of the  magnetic  penetration depth was
calculated within the local (London) approximation
($\lambda\gg\xi$) using the following equation
\cite{Khasanov07_La214}:
\begin{equation}
\frac{\sigma_{sc}(T)}{\sigma_{sc}(0)}=  1+ \frac{8}{\pi-4\varphi_0}
\int_{\varphi_0}^{\pi/4}\int_{\Delta(T,\varphi)}^{\infty}\left(\frac{\partial
f}{\partial E}\right)\frac{E\
dEd\varphi}{\sqrt{E^2-\Delta(T,\varphi)^2}}\ .
 \label{eq:lambda-d}
\end{equation}
$f=[1+\exp(E/k_BT)]^{-1}$ denotes the Fermi function. Here we also
replace the prefactor $8/\pi$ of the integral with
$8/(\pi-4\varphi_0)$ to account for the case when the
superconducting energy gap is developed only on a part of the
Fermi surface (in our case from $\varphi_0$ to $\pi/4$). The
results of this analysis are presented in
Fig.~\ref{fig:sigma_vs_T}a. The monotonic $d-$wave gap as well
as the combined gap represented by the solid blue line in
Fig.~\ref{fig:gap}a  can not describe the experimental
$\sigma_{sc}(T)$. Full consistency between ARPES and $\mu$SR
data is obtained if one assumes a superconducting $d-$wave gap with only carriers in the region $20^{\rm
o}\lesssim\varphi<45^{\rm o}$ contributing to the superfluid.
It should be noted here that the
theoretical $\sigma_{sc}(T)$ curves in Fig.~\ref{fig:sigma_vs_T}
were not fitted, but obtained directly by introducing the
angular dependence of the gap measured in ARPES experiments into
Eq.~(\ref{eq:lambda-d}), describing the $T$ dependence of
the penetration depth within the London approach.

\begin{table}[h]
\caption[~]{\label{Table:results} Transition temperature $T_c$,
zero-temperature in-plane magnetic penetration depth
$\lambda_{ab}^{-2}(0)$, and angular region where the superconducting
$d-$wave gap is observed for OP Bi2201 (studied here), OP Na-CCOC,
and OP La214.
}
\begin{center}
 \vspace{-0.7cm}
\begin{tabular}{lclcccccc}\\ \hline
\hline
 Compound&$T_c$&$\ \ \ \lambda_{ab}^{-2}(0)$&SC gap region\\
&(K)&$\ \ \ (\mu$m$^{-2}$)&\\
\hline
OP Bi2201&35&7.8 &$20^{\rm o}\lesssim\varphi<45^{\rm o}$\ $^{(a)}$ \\
OP Na-CCOC&28&10.0 \ $^{(b)}$&$20^{\rm o}\lesssim\varphi<45^{\rm o}$ \ $^{(c)}$ \\
OP La214&36&15.0 \ $^{(d)}$&$0^{\rm o}\leq\varphi<45^{\rm o}$ \ $^{(e)}$\\
 \hline \hline
\end{tabular}
   \end{center}
$^{(a)}$ Ref.~\onlinecite{Kondo07}, $^{(b)}$ Ref.~\onlinecite{Khasanov07_NaCCOC}, $^{(c)}$ Ref.~\onlinecite{Hanaguri07}, $^{(d)}$ Ref.~\onlinecite{Panagopoulos99}, $^{(e)}$ Ref.~\onlinecite{Shi07}
\end{table}

Next we compare the zero-temperature values of
$\lambda_{ab}^{-2}(0)\propto n_s/m^\ast$ for various OP HTS's
having comparable $T_c$ values and for which the angular
dependence of the superconducting gap was measured
\cite{Kondo07,Hanaguri07,Shi07}. OP La214, which exhibits a
fully developed superconducting gap, has an approximately 50\%
higher value of the superfluid density as compared to both OP Na-CCOC
and OP Bi2201, having the superconducting gap opened only on a limited part of the Fermi surface (see Table~\ref{Table:results}). Assuming that the
supercarrier masses $m^\ast$ are the same for all OP compounds
listed in Table~\ref{Table:results} (in analogy with
$m^\ast\simeq3-4m_e$ reported for La214 and
YBa$_2$Cu$_3$O$_{7-\delta}$ families of HTS's \cite{Padilla05}),
the difference in the values of $\lambda_{ab}^{-2}(0)$ can be naturally 
explained by the different number of carriers condensed into the
superfluid. In the case of OP Bi2201 and OP Na-CCOC, $n_s$ is
strongly reduced because of the fraction of the states is no more
available for the superconducting condensate due to the
pseudogap.


To conclude, the in-plane magnetic penetration depth
$\lambda_{ab}$ in optimally doped Bi2201 was studied by
means of muon-spin rotation. By comparing the measured
$\lambda_{ab}^{-2}(T)$ with the one calculated theoretically using a
model consistent with ARPES measurement \cite{Kondo07} we found
that the superconducting gap in OP Bi2201 has $d-$wave symmetry,
but only carriers from parts of the Fermi surface close to the
node ($20^{\rm o}\lesssim\varphi\lesssim45^{\rm o}$) contribute to
the superfluid.
This implies that the pseudogap affects the spectral density of the
quasiparticles and, consequently, not all the states 
at the Fermi surface are available to participate in the superconducting
condensate. Our results supports the scenario where the superconducting and pseudogap state are two distinct and competing phenomena.
This statement is also consistent with the fact that the superfluid
density in OP Bi2201 is strongly reduced in comparison with that
in OP La214, where the superconducting gap and coherent
quasiparticles are observed along the whole Fermi surface ($0^{\rm
o}\leq\varphi<45^{\rm o}$) \cite{Shi07}.


This work was performed at the Swiss  Muon Source (S$\mu$S), Paul
Scherrer Institute (PSI, Switzerland). The authors are grateful to
Y.J.~Uemura and R.~Prozorov for stimulating discussions, and
S.~Weyeneth for performing torque experiments. This work was
supported by the K.~Alex M\"uller Foundation and in part by the
Swiss National Science Foundation. Work at the Ames Laboratory was
supported by the Department of Energy - Basic Energy Sciences
under Contract No. DE-AC02-07CH11358.


\begin{thebibliography}{99}
%
\bibitem{LeTacon06}  M.~Le~Tacon {\it et al.},
Nature~Physics {\bf 2}, 537 (2006).
%
\bibitem{Tanaka06} K.~Tanaka {\it et al.},
Science {\bf 314}, 1910 (2006).
%
\bibitem{Kondo07} T.~Kondo {\it et al.},
Phys.~Rev.~Lett. {\bf 98}, 267004 (2007).
%
\bibitem{Lee07} W.S.~Lee {\it et al.},
Nature (London) {\bf 450}, 81 (2007).
%
\bibitem{Guyard08} W.~Guyard {\it et al.},
Phys.~Rev.~B {\bf 77}, 024524 (2008).
%
\bibitem{Hanaguri07} T.~Hanaguri {\it et al.},
Nature~Physics {\bf 3}, 865 (2007).
%
\bibitem{Khasanov07_NaCCOC} R.~Khasanov {\it et al.},
Phys.~Rev.~B {\bf 76}, 094505 (2007).
%
\bibitem{Panagopoulos99} C.~Panagopoulos {\it et al.},
Phys.~Rev.~B {\bf 60}, 14617 (1999).
%
\bibitem{Shi07} M.~Shi {\it et al.},
unpublished, arXiv:0708.2333.
%
\bibitem{Kondo04Kondo05} T.~Kondo {\it et al.},
J.~Electron~Spectrosc.~Relat.~Phenom. {\bf 137--140}, 663 (2004);
%
T.~Kondo {\it et al.},
Phys.~Rev.~B {\bf 72}, 024533 (2005).
%
\bibitem{Brandt03} E.H.~Brandt, Phys.~Rev.~B {\bf 68}, 054506
(2003).
%
\bibitem{Khasanov06a} R.~Khasanov {\it et al.},
Phys.~Rev.~B {\bf 73}, 214528 (2006).
%
\bibitem{Musolino03} N.~Musolino {\it et al.},
Physica~C {\bf 399}, 1 (2003).
%
\bibitem{Lee93} S.L.~Lee {\it et al.},
Phys.~Rev.~Lett. {\bf 71}, 3862 (1993).
%
\bibitem{Aegerter98} C.M.~Aegerter {\it et al.},
Phys.~Rev.~B {\bf 57}, 1253 (1998).
%
\bibitem{Russo07} P.L.~Russo {\it et al.},
Phys.~Rev.~B {\bf 75}, 054511 (2007).
%
\bibitem{Weyeneth08} S.~Weyeneth, under preparation.
%
\bibitem{Kawamata99} S.~Kawamata {\it et al.},
J.~Low.~Temp.~Phys. {\bf 117}, 891 (1999).
%
\bibitem{Wang03} Y.~Wang {\it et al.},
Science {\bf 299}, 86 (2003).
%
\bibitem{Amin00} M.H.S.~Amin, M.~Franz, and I.~Affleck, Phys.~Rev.~Lett. {\bf 84}, 5864 (2000).
%
\bibitem{Vekhter99} I.~Vekhter, J.P.~Carbotte, and E.J.~Nicol, Phys.~Rev.~B {\bf 59}, 1417 (1999).
%
\bibitem{Carrington03} A.~Carrington and F.~Manzano, Physica~C {\bf 385}, 205 (2003).
%
\bibitem{Khasanov07_La214} R.~Khasanov {\it et al.},
Phys.~Rev.~Lett. {\bf 98}, 057007 (2007).
%
\bibitem{Padilla05} W.J.~Padilla {\it et al.},
Phys.~Rev.~B {\bf 72}, 060511 (2005).


\end{thebibliography}
\end{document}